\documentclass[fleqn,10pt]{wlscirep}
\usepackage[utf8]{inputenc}
\usepackage[T1]{fontenc}
\usepackage{makecell}
\usepackage{caption}
\title{Implementation of Reservoir Computing Using Coupled Microelectromechanical Drum Resonators via Sideband-Pumped Phonon–Cavity Dynamics}
\author[1]{Theresa Farah}
\author[1]{Loïc Flis}
\author[1]{Pierre Laly}
\author[2]{Guo-En Chang}
\author[3]{Jun-Yu Ou}
\author[3]{Yoshishige Tsuchiya}
\author[1]{Yan Pennec}
\author[1]{Bahram Djafari-Rouhani}
\author[1,*]{Xin Zhou}
\affil[1]{CNRS, University of Lille, Centrale Lille, Univ. Polytechnique Hauts-de-France, UMR 8520 IEMN, F-59000 Lille, France}
\affil[2]{Department of Microelectronics, National Yang Ming Chiao Tung University, Hsinchu City 300093, Taiwan.}
\affil[3]{University of Southampton, Southampton SO17 1BJ, United Kingdom}
\affil[*]{xin.zhou@cnrs.fr}


\begin{abstract}
Reservoir computing is a bio-inspired machine learning paradigm that exploits the intrinsic dynamics of nonlinear systems with fading memory for efficient temporal information processing. Microelectromechanical resonators offer a promising platform for reservoir computing as they inherently possess the requisite nonlinear and temporal properties while also facilitating the integration of sensing and computing within a single platform. In this work, we experimentally demonstrate a physical reservoir computing platform based on two capacitively coupled drum resonators, operating in the MHz frequency regime. Taking advantage of the concept of phonon-cavity electromechanics, a pump tone is applied at the sideband of the phonon cavity while probing one of the coupled modes, analogous to optomechanical systems, thereby creating nonlinear dynamics in energy transfer between the two resonators. Reservoir computing is implemented by exploiting the nonlinear response generated through pump amplitude modulation in combination with a time-delay feedback loop, and the performance is evaluated using both parity and Normalized Auto-Regressive Moving Average benchmarks. This work demonstrates a compact microelectromechanical platform for integrated sensing and reservoir computing and shows that the sideband pumping scheme provides a pathway for extending conventional single-resonator reservoir computing toward multimode architectures.
\end{abstract}
\begin{document}

\flushbottom
\maketitle
%
%
\thispagestyle{empty}
\section{\label{sec:intro} Introduction}
Reservoir Computing is a bio-inspired machine learning paradigm designed for processing complex time-series data \cite{RC_overview,jaeger, maass}. It has been successfully applied to a diverse range of temporal tasks, including chaotic time-series prediction, speech recognition, and signal classification. As a simplified framework derived from Recurrent Neural Networks (RNNs) \cite{graves,werbos}, reservoir computing fixes the internal weights of the reservoir and only requires the output layer weights to be trained by using simple linear regression. It therefore significantly reduces computational costs while circumventing the vanishing gradient problems typical of traditional RNN training. A physical system used to implement a reservoir must exhibit two key properties: nonlinearity and fading memory \cite{samarasinghe, Feedbback}. Nonlinearity is required to project input signals into a high-dimensional state space, enabling complex data to become linearly separable at the readout layer. Fading memory is essential for processing temporal sequences, as it establishes a connection between the current reservoir state and recent past inputs. To date, reservoir computing has been demonstrated across a wide range of hardware platforms, including photonic and optoelectronic systems \cite{optoelectronic_RC_1,optoelectronic_RC_2,optics_RC}, robotics \cite{mechanics_RC_2, mechanics_RC_1}, memristor arrays \cite{memristors_RC}, spintronic \cite{torrejon2017neuromorphic, spin_waves_2,RC_spin_waves}. More recently, reservoir computing has been further extended beyond classical systems to quantum platforms, including quantum oscillators and nonlinear oscillators coupled to quantum systems \cite{dudas2023quantum, govia2021quantum}.\\

%

Micro-Electro-Mechanical Systems (MEMS), which enable mechanical degrees of freedom to be coupled with electrical signals, constitute a promising platform for reservoir computing \cite{Sylvestre, coulombe, guo2024mems, Sun2021,shima2025improved}. MEMS naturally possess the two essential properties required for reservoir computing: intrinsic nonlinearity arising from geometric and material constraints (such as the Duffing nonlinearity) effects, and fading memory facilitated by the finite decay time of their mechanical resonances \cite{bachtold2022mesoscopic,cattiaux,maillet2018measuring}. In addition, MEMS devices can be designed as scalable architectures and are easily integrated with modern CMOS (Complementary Metal–Oxide–Semiconductor) platforms \cite{fedder2015resonant}. Beyond these advantages, one of the most compelling motivations for developing MEMS-based reservoir computing is its potential to combine sensing and computing within a single platform \cite{nikfarjamenergy, bai2025mems}. Such integration will overcome the inefficiencies of conventional architectures, where the physical separation of sensing and processing units leads to additional energy consumption and latency due to intensive data transfer. To date, MEMS-based physical reservoir computing has been experimentally implemented primarily using single electromechanical resonators, exploiting driving-force modulation in the Duffing nonlinear regime \cite{Sylvestre} and stiffness modulation techniques \cite{guo2024mems}. Only a limited number of studies have explored reservoir computing in coupled MEMS systems. These efforts include numerical investigations of reservoir performance in coupled resonators \cite{zheng2021enhancing} and experimental studies evaluating memory capacity in coupled triple-resonator configurations \cite{shima2025improved}. However, they rely on mechanical coupling and require the resonators to have closely matched resonance frequencies, which places stringent demands on nanofabrication and poses challenges for scaling to higher integration densities.\\
%

In this work, we experimentally demonstrate a novel reservoir computing scheme based on coupled microelectromechanical resonators by exploiting concepts from phonon-cavity electromechanics \cite{mahboob2012phonon,sun2016correlated}. The device employed is a double-drum electromechanical system consisting of two capacitively coupled membrane resonators operating in the MHz frequency regime \cite{xin_zhou,alok}. Drawing an analogy to two-tone operation in optomechanical systems \cite{kumar2023microwave}, we apply a pump tone at the sideband of the phonon-cavity while probing one of the coupled modes, thereby inducing nonlinear dynamics in the phonon transfer between the two vibrational modes \cite{alok}. By modulating the pump amplitude and implementing a time-delay feedback loop, we demonstrate reservoir computing by probing one of the coupled drum resonators and evaluating both parity and Normalized Auto-Regressive Moving Average (NARMA) benchmarks. The unique device design and pump amplitude modulation scheme point toward a compact MEMS platform for co-integrated sensing and reservoir computing. Moreover, the proposed scheme does not require the resonance frequencies to fall within a narrowly distributed range and is not limited to the specific double-drum design. It can be easily extended to other mechanically coupled resonator arrays and multimode optomechanical systems, in both the classical and quantum regimes.

\section{Results}
\subsection{Double-drum resonator and setup for reservoir computing}
\label{sec:dev}
\begin{figure}
    \centering
    \includegraphics[width=1\linewidth]{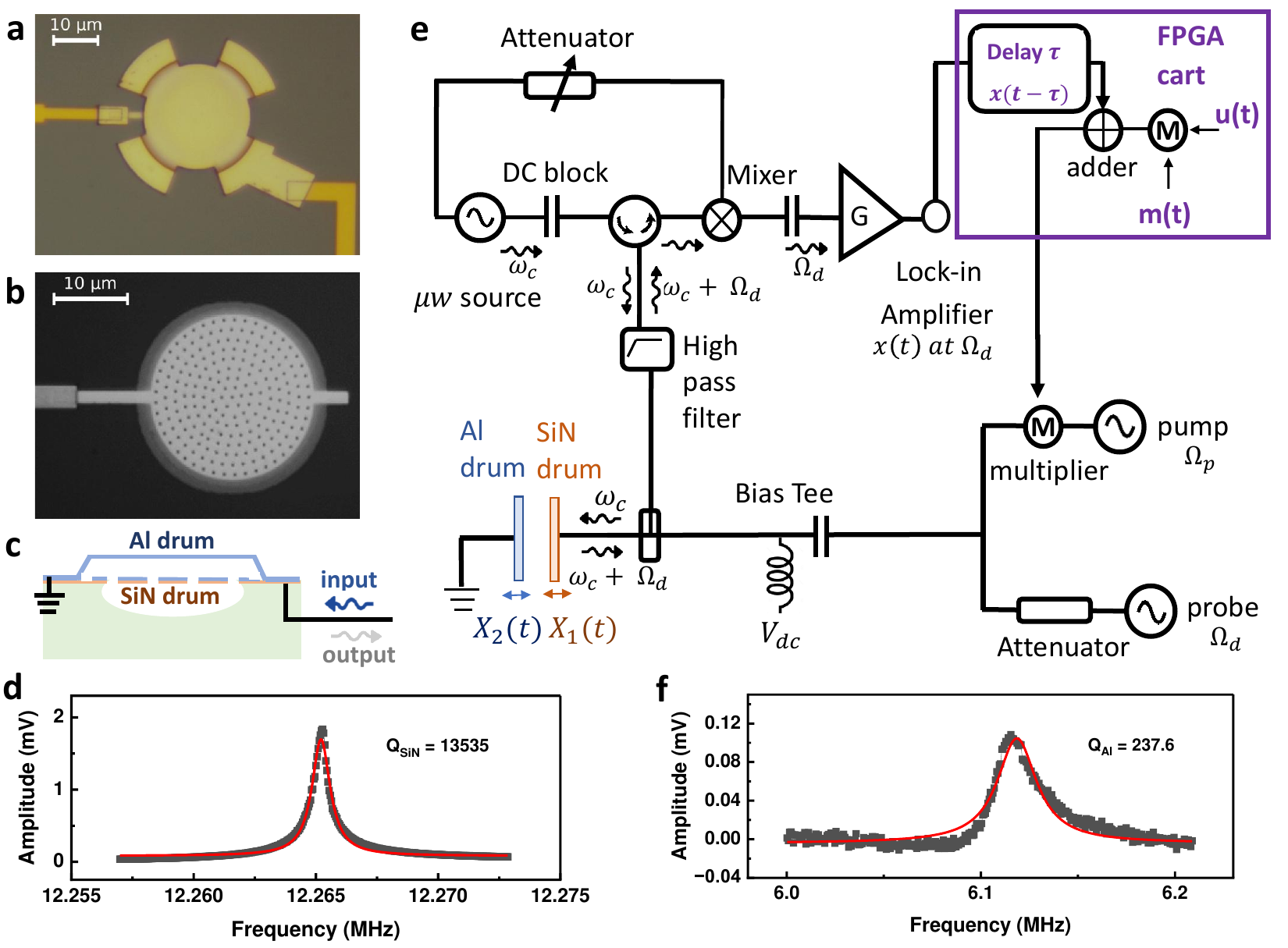}
    \caption{Experiment setup. \textbf{a} Optical image of the Al drum resonator which is suspended over the SiN circular drum by the means of 4 support feet. \textbf{b} Optical image of the bottom SiN membrane resonator, covered with an Al thin film. There is no physical connection between the two drums. \textbf{c} Cross sectional view of the electromechanical system. \textbf{d} and \textbf{f} The mechanical responses of the SiN and Al drum resonators are measured under a $DC$ bias of $V_{dc}$ = 4 V, with $AC$ drive of $V_{d}$ = 6 mV and 80 mV, respectively, each attenuated by 20 dB. The red lines represent the Lorentz fitting result. \textbf{e} Schematic diagram of the measurement setup. The mechanical displacement is read out using a lock-in amplifier via microwave optomechanical interferometry. The output of the FPGA is used to modulate the pump signal through a multiplier.}
    \label{fig:setup}
\end{figure}
%
The microelectromechanical system used to perform reservoir computing in this work is a double-drum resonator, consisting of two suspended membrane electromechanical resonators, as shown in Fig. \ref{fig:setup}a-c. A suspended aluminum (Al) membrane serves as a capacitively coupled top gate to the underlying silicon nitride (SiN) drum microelectromechanical resonator \cite{xin_zhou, alok}. In order to excite mechanical vibrations, electrostatic forces are generated by applying a combination of a $DC$ voltage $V_{dc}$ and an $AC$ voltage $V_{ac}$ at a drive frequency $\Omega_d$ close to the mechanical resonance frequency $\Omega_m$ of the target mechanical resonator. We exploit the microwave optomechanical interferometry to simultaneously readout the vibrations of both drums, as shown in Fig. \ref{fig:setup}e. Microwave photons at a frequency $\omega_c$ = 6 GHz are shined to the double-drum resonator through a 50 Ohm transmission line. The reflected microwave signal, carrying information of the mechanical vibrations at frequency $\omega_c+\Omega_d$, is readout by using a lock-in amplifier after frequency down-conversion \cite{xin_zhou}. Mechanical mode vibrations are therefore measured in volts in our experiment. Using this readout scheme, the double-drum resonator is initially characterized in its linear operating regime by applying probe tones around $\Omega_m$ with small $AC$ amplitudes $V_{d}$, as shown in Fig. \ref{fig:setup}d-f. The SiN drum has a resonance frequency of $\Omega_{SiN}/(2\pi) \approx$ 12.265 MHz and a quality factor of $Q_{SiN}\approx$ 1.3 $\times 10^{4}$. The Al drum resonates at the frequency $\Omega_{Al}/(2\pi)\approx$ 6.13 MHz with a quality factor $Q_{Al}\approx$ 237. All measurements are carried out at room temperature and under vacuum conditions, $\approx6\times 10^{-6}$ mbar.\\     

The reservoir used for computing is constructed from virtual nodes based on a single double-drum resonator acting as the physical node. The creation of virtual nodes is inspired by the conventional time domain multiplexing method implemented in previous works \cite{Feedbback, Sylvestre}. A standard time-delay loop is implemented using a field-programmable gate array (FPGA), which adds a time delay $\tau$ to the detected mechanical vibration $x(t)$ from the output of the nonlinear element and feeds it back to the input of the resonator in the form of driving forces. The virtual nodes are created by mapping a predefined mask $m(t)$ onto the input data $u(t)$. The masked input is then multiplexed with the delayed output of the physical element, $x(t-\tau)$, and combined through a multiplier to generate additional modulations on the driving forces, acting on the double-drum resonator. 
\subsection{Nonlinear behavior generated by sideband pumping in phonon-cavity electromechanics}
\label{sec:sideband_pumping}
\begin{figure*}[ht]
\centering
  \includegraphics[width=1\textwidth]{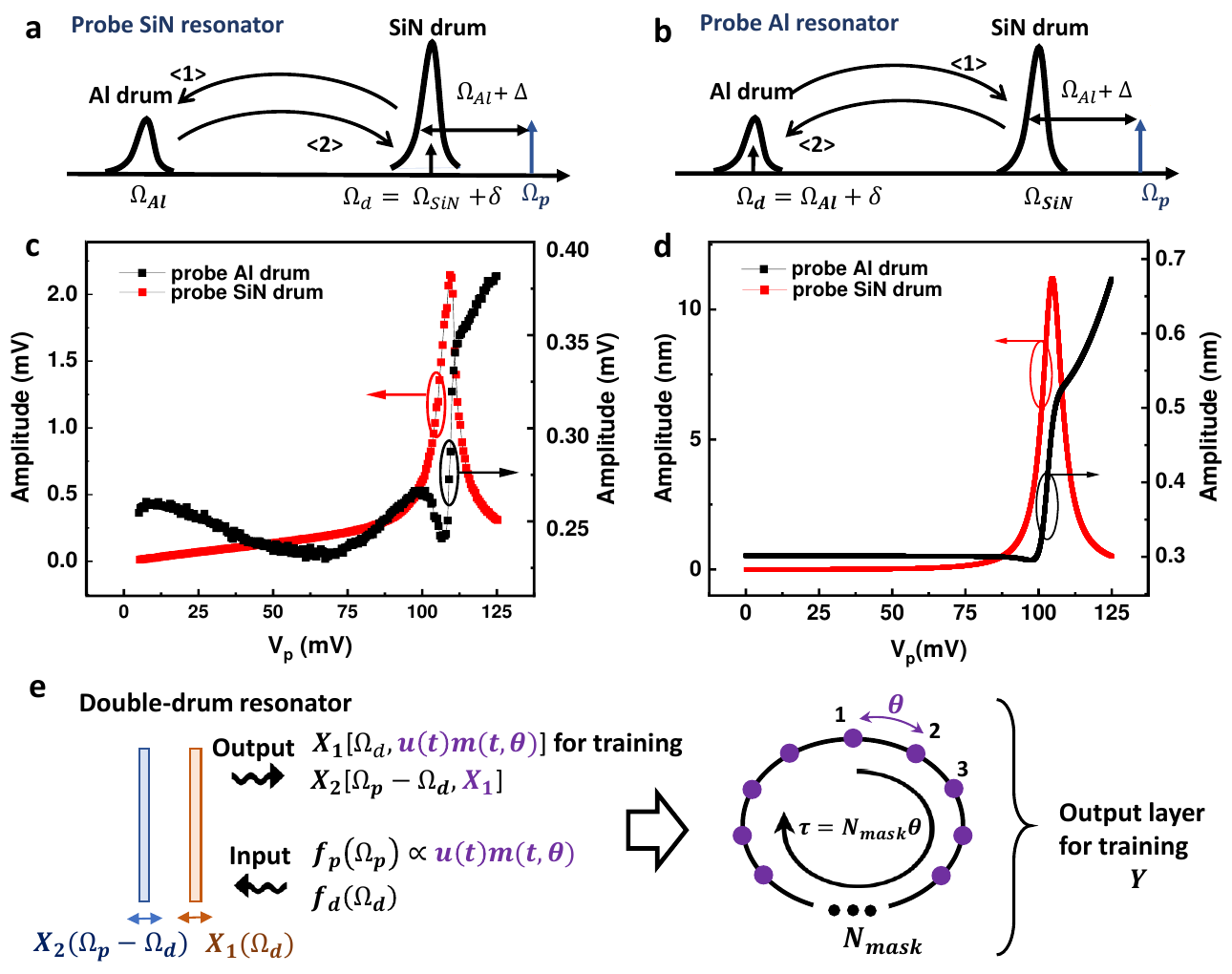}
 \caption{Diagram of the blue sideband pumping scheme while probing \textbf{a}: The phonon-cavity SiN drum and \textbf{b}: The Al membrane resonator, respectively. \textbf{c} The measured mechanical responses as a function of the pump amplitude. The background contribution has been subtracted using the measured minimum amplitude as a reference. The black (red) curve is measured by probing Al (SiN) drum at $\Omega_d/(2\pi)$ = 6.12 MHz ( = 12.25 MHz) with amplitude $V_d$ = 80 mV (6 mV) attenuated by 20 dB, with $\Omega_p/(2\pi)$ = 18.38 MHz and $V_{dc}$ = 4 V. \textbf{d} Analytical calculation results of the mechanical amplitude as a function of the pump amplitude in both probing cases, by using experimental parameters. \textbf{e} Schematic of the concept for implementing reservoir computing in coupled drum resonators. Phonons carrying the virtual nodes are generated simultaneously in both vibrating drum modes.}
  \label{fig:sideband pumping}
\end{figure*}

%
The key concept of phonon-cavity electromechanics is to manipulate coherent energy transfer between two coupled mechanical vibration modes, $\Omega_1$ and $\Omega_2$,  by selecting the mode having the higher resonance frequency ($\Omega_1 > \Omega_2$) and pumping it at its sideband $\Omega_1\pm \Omega_2$ \cite{mahboob2012phonon, sun2016correlated, alok, xu2024imaging}. It inherits rich physics from optomechanics, enabling cooling of mechanical vibration modes, amplification of phonon occupancy, and the creation of controllable interference effects \cite{zhou2021electric}. In this double-drum resonator, we choose the SiN drum as the phonon-cavity mode and pump it at its blue sideband, at the frequency $\Omega_{p} = \Omega_1+ \Omega_2 + \Delta= \Omega_{Al} + \Omega_{SiN} + \Delta$. Here the $\Delta$ is defined as small frequency detuning. Additional to the pump tone, a second tone with small amplitudes is introduced to  probe one of the mode with a small frequency detuning $\delta$ from the resonance, either $\Omega_d\approx\Omega_1+\delta$ or $\Omega_d\approx\Omega_2+\delta$, placing the double-drum under a two-tone driving scheme, as shown in Fig. \ref{fig:sideband pumping}a and b. Because of frequency mixing effects, the probe and pump tones generate vibration phonons in the unprobed mode, as indicated by the process $<1>$. These phonons are subsequently projected back onto the probed mode by the pump, where they interfere with the probe tone, as marked by the process $<2>$. \\

The two-tone dynamics of the double-drum resonator are described by the following coupled equations of motion:
\begin{equation}
    \begin{aligned} 
    \ddot{X_{1}}+\gamma_1 \dot{X_1}+\Omega_{1}^2 X_1 &=\frac{V_{ac}V_{dc}}{m_1d}C_{go}\left[ 1-2\frac{({X_2} -{X_1})}{d}\right], \\
    \ddot{X_2}+\gamma_2 \dot{X_2}+\Omega_{2}^2 X_2 &=\frac{V_{ac}V_{dc}}{m_2d}C_{go}\left[- 1+2\frac{({X_2} -{X_1})}{d}\right], \\
    \end{aligned} 
    \label{eq:motion}
\end{equation}
where index 1 and 2 refer to each of the SiN and Al membrane respectively. The $m_{1,2}$ denotes the effective mass of the resonator, $d$ the distance between two drums, $X_{1,2}$ the mechanical displacement, $C_{g0}$ coupling capacitance between two drums at $X_{1,2}$ = 0, and $\gamma$ the damping rate. Within the capacitive coupling scheme, the electrostatic force acting on the resonator arises from the term $V_{dc}V_{ac}\partial C_{g}(x_{1,2})/\partial x_{1,2}$ \cite{xin_zhou}. In the two-tone driving scheme, $V_{ac}$ is given by $ V_p \cos(\Omega_pt) +V_d \cos(\Omega_dt)$. The probed resonator is driven by two forces, one coming from the driving tone $V_d \cos(\Omega_dt)$ and the other arising from the pump tone induced term $V_p \cos(\Omega_pt)\cdot X_{1,2}$, in a capacitive coupling scheme. The latter term, referred to as the cavity force \cite{zhou2021electric, alok}, enables the pump mediated coupling between the two resonators. Therefore, this sideband pumping technique enables coherent energy transfer between two capacitively coupled resonators even when their resonance frequencies are significantly different. In the case of probing the SiN drum resonator, $\Omega_d$ = $\Omega_{1}+\delta$ = $\Omega_{SiN}+\delta$, the displacement of the probed resonator can be obtained by solving the Eq.\ref{eq:motion}. To do so, we re-write variables in complex form, such as $X_{1}(t) = \frac{x_{1}(t)}{2}e^{-i\Omega_{d}t}+c.c.$, and $x_1$ arrives

\begin{equation}
\begin{aligned}
    x_{1} &= \frac{f_d}{2m_1\Omega_1}
        \frac{1}{
            \frac{1}{\chi_1}
            +g^2\chi_2 
        }, \\
    x_{2} &= \frac{f_p^*}{2 m_2 \Omega_2}\frac{x_1}{d}\chi_2,
\end{aligned}
\label{eq:x1x2}
\end{equation} 
by looking for the solution of $x_1$ around the probe frequency $\Omega_d$. Here, $f_d = \frac{C_{g0}V_{dc}\mu_d}{d}$ and $f_p = \frac{C_{g0}V_{dc}\mu_p}{d}$ are the amplitudes of the probe and pump forces respectively, the parameter $\mu_d$ and $\mu_p$ represent the complex amplitude of the probe and pump tone respectively, and the $c.c.$ denotes the corresponding complex conjugate term. The $\chi_1 = \frac{1}{-\delta -i\frac{\gamma_1}{2}}$ and $\chi_2 = \frac{1}{\delta-\Delta -i\frac{\gamma_2}{2}}$ are the mechanical susceptibilities of the drum resonators when probing the SiN phonon-cavity. We define an effective coupling strength induced by the pump tone and the phonon population as, $g^2$ = $\frac{\vert f_p^2 \vert}{4m_1 m_2 d^2\Omega_1\Omega_2}$. In the expression of Eq.\ref{eq:x1x2}, the $g^2$ acts to modulate the effective susceptibility of the SiN resonator. The solutions of the coupled motion equations shown in the Eq.\ref{eq:x1x2} clearly indicate that the phonons in the unprobed mode, here the $x_2$, are generated by both the probe and pump tones through frequency conversion between the $f_d$ and the $f_p$. The entire derivation process has been reported in our previous work \cite{alok}. The calculation results corresponding to the case of probing the Al drum are provided in the Supporting Information (SI). From the displacement expressions of both drums, $x_1$ and $x_2$, as given in Eq.\ref{eq:x1x2}, the blue sideband pump brings nonlinear amplifications of the phonons for the probed mode through creating frequency dependent constructive interference. Such interference arises from pump induced coherent phonon cycling between the two coupled modes, and the effective constructive interferences windows depend on the frequency detunings $\delta$ and $\Delta$ relative to the mechanical damping rates $\gamma_{1}$ and $\gamma_{2}$. The pump tone functions as a phonon bus, transferring the energy from one mode to the other, in the coupled resonators. \\

To take advantage of this pump induced nonlinear behavior for reservoir computing, we exploit the fact that the sideband pump amplitude has an effect on shifting $\Omega_{SiN}$, so called optical spring effect in optomechanics \cite{alok, zhou2021electric}. More details can be found in SI part. This effect arises from the fact that the sideband pump amplitude modifies the effective susceptibility of the coupled mechanical resonators. It is worth emphasizing that this pump induced effect on $\Omega_{Al}$ can be neglected in this double-drum system. Because the Al drum has a relatively large bandwidth (damping rate), $\gamma_{Al} \approx 2\pi \times$ 25.9 kHz, small variations in the resonance frequency are not easily observed. To do so, we set the pump frequency to $\Omega_p =\Omega_{SiN}+\Omega_{Al}$, where both resonances $\Omega_{SiN}$ and $\Omega_{Al}$ are obtained by measuring mechanical responses at the sideband pump amplitude $V_{pc}$  = 105 mV. We then set the probe tone frequency to either $\Omega_d=\Omega_{SiN}$ or $\Omega_d=\Omega_{Al}$ + a few kHz, and measure its amplitude as a function of $V_p$. The detected probe signals exhibit clear nonlinear constructive interference induced by variations in the pump amplitude, as shown in  Fig. \ref{fig:sideband pumping}c. When the pump amplitude deviates far from the calibrated value $V_{pc}$ = 105 mV, the interference between the probe tone and the phonons projected back from the unprobed mode becomes less effective. This is due to shifts in $\Omega_{SiN}$, which induce frequency de tuning values of the $\delta$ and $\Delta$, appearing in the susceptibilities $\chi_1$ and $\chi_2$ in Eq.\ref{eq:x1x2}. These detuning parameters, relative to the linewidth of the mechanical resonators, define the effective interference window in the frequency space. These interference features have been experimentally measured and modeled in our previous work \cite{alok}. Because the two drum resonators have very different damping rates, the pump induced nonlinear behaviors are quite different when probing the cavity mode versus the Al drum. The bandwidth $\gamma_{SiN} \approx 2\pi \times$ 944 Hz is much smaller than the resonance-frequency shift induced by variations in the pump amplitude (see SI). Consequently, when the probe tone is initially biased at $\Omega_{SiN}$ corresponding to $V_{pc}$ = 105 mV, both increases and decreases in the pump amplitude can shift $\Omega_{SiN}$ over a frequency rang of 7 kHz, making the probe tone to move out of resonance. However, the probe tone can still be amplified through constructive interference, since this frequency shift lies within $\gamma_{Al}$. When probing the Al drum, although the probe tone always remains within the bandwidth, clear constructive interference can be observed only when the resonance shift relative to the value calibrated at $V_{pc}$ lies within $\gamma_{SiN}$. We also noticed that the detected amplitudes of Al drum exhibiting small fluctuations for $V_p <$ 75 mV. It may be induced by fluctuations in the background noise carried by the pump tone. Fig. \ref{fig:sideband pumping}d presents the calculation results, based on Eq.\ref{eq:x1x2}, including the pump induced shift of $\Omega_{SiN}$, which are consistent with the experimental observations. These results demonstrate that pump amplitude modulation in the two-tone driving scheme provides a controllable way of inducing nonlinear behavior in the probe tone, in coupled resonators. In the following part, we exploit this nonlinear dynamics for reservoir computing. 

\subsection{Reservoir computing implementation in the phonon-cavity scheme}
In the experiment setting, each input is held for a delay time $\tau$, during which a random binary mask consisting of $N_{mask}=400$ random values ranging between 0 and 1 is applied with a sampling period of $\theta$, yielding $\tau$ = $N_{mask}\theta$. The parameter $N_{mask}$ defines the numbers of the virtual nodes, which correspond to $N_{mask}$ equidistant points separated in time by $\theta$. The delay feedback loop provides the reservoir states with fading memory by coupling the current mechanical response $x(t)$ to its delayed response $x(t-\tau)$, thereby creating a recurrent loop in the virtual neural network. As shown in the Fig.  \ref{fig:setup}e, both the masked input $u(t)m(t)$ and the delayed mechanical responses x(t-$\tau$) modulate the pump amplitude by combining the FPGA output with the $AC$ signal through a multiplier. This modulation can be described by expressions of
\begin{equation}
    V_{p}(t) \propto V_{p0}[\alpha_{1}u(t)m(t) + gl\cdot\alpha_{2}[x(t-\tau)-x_{0}] + offset].
    \label{eq:modulation}
\end{equation}
The values of the constants $\alpha_1$, $\alpha_2$ and $offset$ are fixed by programmed MATLAB codes to avoid the saturation of the FPGA card and the multiplier. The $gl$ is an amplification factor set by the lock-in amplifier to ensure a sufficient signal-to-noise ratio when the delayed signal is processed in the FPGA. The typical value of the $gl$ is $\sim$100 for signals detected from the SiN mode and $\sim$400 for or those from the Al mode. Here, the $x_0$ denotes the preset offset in the measurement. These values define the modulation range of the pump amplitude and ensure linear operation of the modulation, while enabling the nonlinearity arising from pump-induced interference required for reservoir computing. In the phonon-cavity two-tone scheme, the masked input data $u(t)m(t,\theta)$ are applied to the pump force $f_p$, which transfers the information to the mechanical displacements of the two coupled drum resonators. The virtual nodes are mapped onto both drum vibrating modes via nonlinear dynamics induced by pump-mediated interactions between the two mechanical modes, thereby enriching the system dynamics and introducing an additional physical degree of freedom that provides spatial dimensionality to the reservoir. Figure \ref{fig:sideband pumping}e present the schematic of this concept. In principle, based on Eq.~\ref{eq:motion}, the displacement of either drum resonator could be used for the training process. However, in our experiment we choose to read out the probed resonator in order to have the better signal-to-noise ratio. \\

\subsection{The Parity Benchmark}
The performance of the system was first evaluated using the parity benchmark, which tests the short-term memory capacity of the reservoir and its ability to perform nonlinear mixing of past input states.  This benchmark consists of a sequence of random binary input values $u(t)$, taking either $+1$ or $-1$, each held for a duration $\tau$. The target $n^{th}$ order parity benchmark output is computed as follows : 
\begin{equation}
    P_n(t) = \prod_{i=0}^{n-1}u[t-i\cdot\tau].
\end{equation}
The objective is to train the output weights of the system to provide an output which can be as close as possible to the target output $P_n$. For $n>1$, the parity function is nonlinear separable and the task requires the reservoir
to store information about a nonlinear transform of previous inputs \cite{Sylvestre}. \\

\begin{figure*}
  \centering
  \includegraphics[width=1\textwidth]{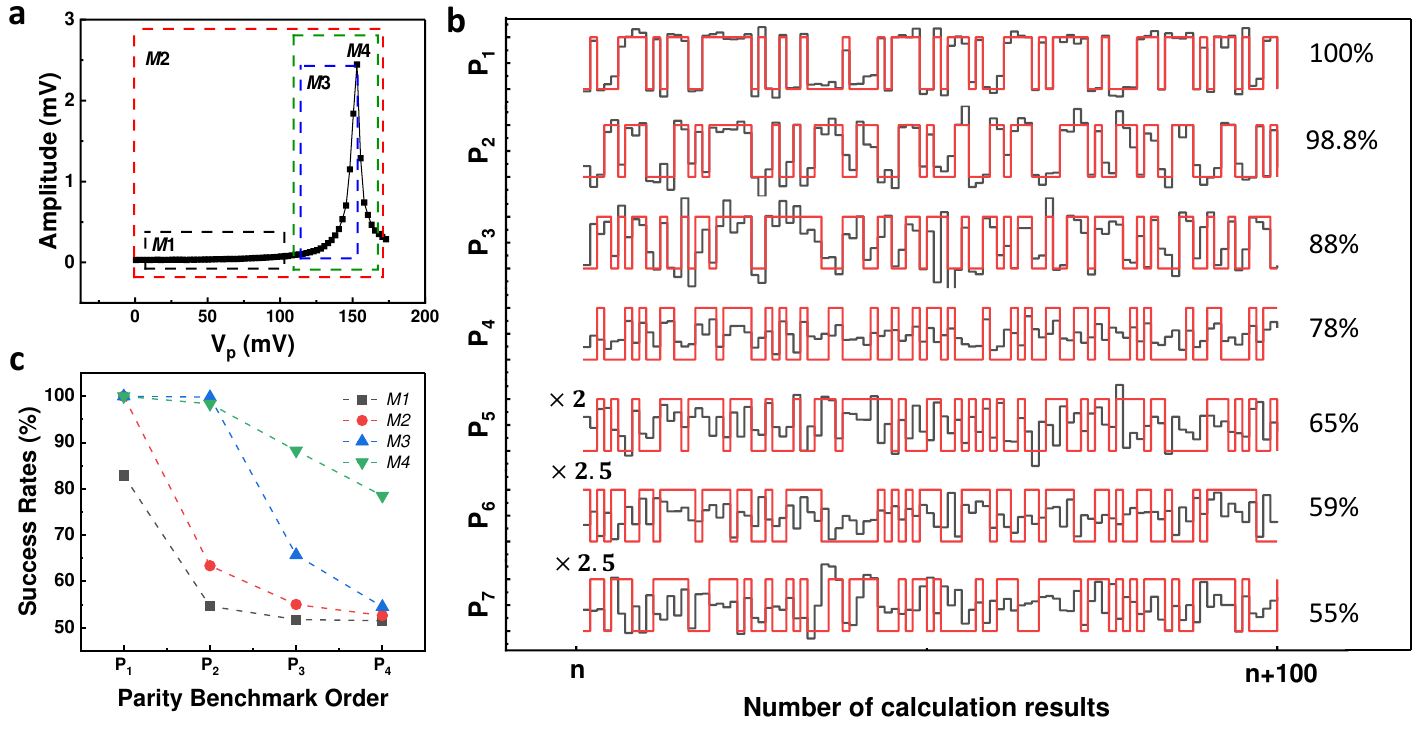}
  \caption{Reservoir computing based on the sideband pumping mechanism while probing the SiN drum with $\Omega_d/(2\pi) = 12.26$ MHz, $V_{d} = 6$ mV attenuated by 20 dB, $\Omega_p/(2\pi) = 18.38$ MHz and $V_{dc} = 4$ V. \textbf{a} The detected mechanical displacement of the probed SiN membrane as a function of the amplitude of the blue pump signal. The curve is divided into 4 modulation windows $M1$, $M2$, $M3$, and $M4$ showcasing different profile variations depending on the interval of the blue pump voltage. \textbf{b} Comparison between the predicted output by our reservoir computing scheme (black lines) and the target ideal output of the parity benchmark (red lines) for the first 7 orders of the parity benchmark in the case of the modulation window $M4$. \textbf{c} Success rates obtained for the first 4 orders of the parity benchmark corresponding to each of the 4 modulation windows.}
  \label{fig:parity_ProbSiN}
\end{figure*}

\textbf{Probing the SiN phonon-cavity} We first implement the nonlinear amplification scheme by probing the phonon cavity with a weak probe tone while simultaneously applying a blue-sideband pump, as shown in Fig. \ref{fig:parity_ProbSiN}a. 
Based on this measurement results and Eq.\ref{eq:modulation}, we select the modulation range of the applied pump force by adjusting the setup parameters $\alpha_1$, $\alpha_2$, $offset$ and $V_{p0}$. This enables us to perform separate reservoir-computing measurements under different nonlinear operating regimes. The $M1$ is the window characterized by a relatively linear variation of the response as a function of the pumping force, $M2$ is the whole curve with all the linear and nonlinear profiles, $M3$ is where the nonlinear amplification occurs due to the constructive interferences. The $M4$ corresponds to a combination of the amplified response and a subsequent drop induced by the $\Omega_{SiN}$ shifting out of the effective interference window. \\

Figure \ref{fig:parity_ProbSiN}b shows one example of the comparison between the target values $P_n(t)$ and the predicted values by the neural network, which are obtained in the modulation window $M4$, with $\theta = 100$ $\mu s$ and a period of $\tau = N_{mask} \theta = 40$ ms. We set the constants $\alpha_1$, $\alpha_2$ and $offset$ to 0.55, 0.34 and 0.69, respectively, which are obtained from an optimization process designed to achieve the best success rates while avoiding saturation of the FPGA card. The success rates show near-perfect results for the first-order $P_1$ to the third-order $P_3$ of the parity benchmark, then the curve becomes more noisy for the $P_7$. This decrease in the success rate can be attributed to the lack of higher memory capacity in our device. \\

Figure \ref{fig:parity_ProbSiN}c presents the reservoir computing results for the different modulation windows described above. For the modulation windows of M2 and M3, the success rates decrease rapidly after P$_2$, dropping below 70$\%$. The best results are obtained for $M4$ where the characterization curve is strongly nonlinear and where the influence of the interferences are the most dominant. Consequently, the $M3$ modulation window yields the second-best performance due to its nonlinear response profile. However, as for the $M1$ window, very poor results are achieved reaching approximately 50 \% starting from $P_2$. This is due to the absence of nonlinearity in this modulation window and low signal to noise ratio. Lastly, the window $M2$ in which the force is modulated between 0.5 mV and 175 mV shows also poor results reaching around 60 \% at the second-order. In this modulation range, the linear plateau spans a much wider interval than the nonlinear region, reducing the likelihood that the modulated pump amplitude enters the $V_p$ range that generates nonlinear behavior. Consequently, the success rates decrease rapidly, indicating poor reservoir computing performance for this pump tone modulation window. \\

\begin{figure*}
  \centering
  \includegraphics[width=1.0\linewidth]{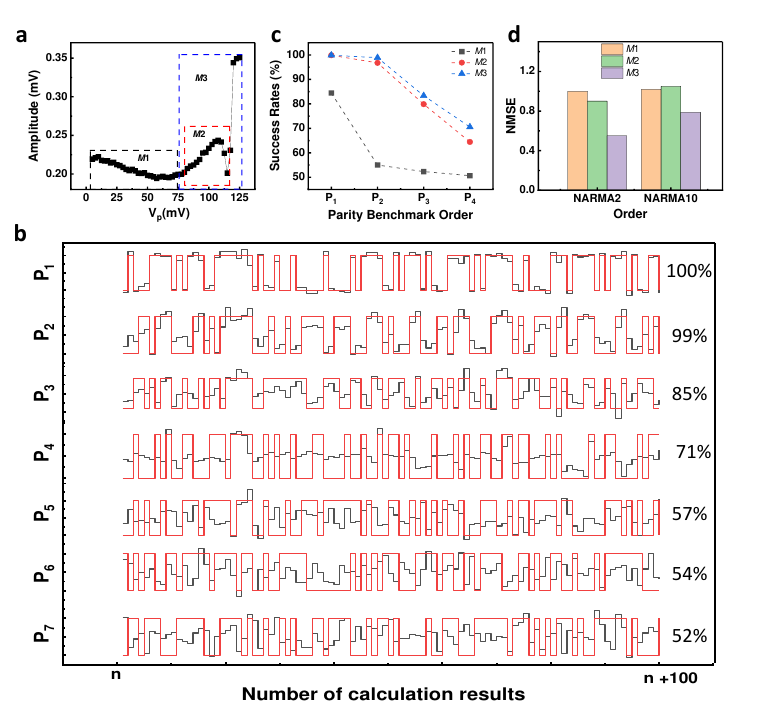}
  \caption{ Reservoir computing based on the sideband pumping mechanism while probing the Al drum with $\Omega_d/(2\pi) = 6.12$ MHz, $V_{d} = 80$ mV (before a 20 dB attenuation), $\Omega_p/(2\pi) = 18.38$ MHz and $V_{dc} = 4 $ V.
  \textbf{a} The detected mechanical displacements of the probed Al membrane as a function of the variation of the amplitude of the blue pump signal. The curve is divided into three modulation windows of the pump amplitudes the $M1$, $M2$ and $M3$ where the detected probe tone exhibits different variations. \textbf{b} Comparison between the predicted output by our reservoir computing scheme (black lines) and the target ideal output of the parity benchmark (red lines) for the first 7 orders of the parity benchmark in the case of the modulation window M3 shown in the \textbf{a}. \textbf{c} Success rates for the first four orders of the parity benchmark (from $P_1$ to $P_4$) for the three different modulation windows.
  \textbf{d} The values of the obtained NMSE in the case of testing NARMA2 and NARMA10 tasks, corresponding to each pump force modulation window shown in the \textbf{a}.}
  \label{fig:parity_ProbAl}
\end{figure*}
\textbf{Probing the Al drum resonator} We use the same probing method described above for the phonon cavity to obtain the nonlinear amplification curve of the probe tone around the $\Omega_{Al}$, as shown in Fig. ~\ref{fig:parity_ProbAl}a. This curve was subdivided into 3 different modulation windows: $M1$ marked by a linear plateau fluctuating around a minimum value of 0.2 mV, $M2$ where the curve exhibits a nonlinear amplification up to a maximum of 0.24 mV accompanied by a fast drop and $M3$ that combines the $M2$ window with the sudden jump to a displacement value of 0.34 mV. The reservoir computing studies with these 3 modulation windows were conducted with values of the force modulation parameters $\alpha_1$, $\alpha_2$ and offset respectively 0.56, 0.33 and 0.71, in order to have the best success rate. Promising success rates were obtained in the cases of the windows $M2$ and $M3$ due to the nonlinear variation of the characterization curve in each window. Near perfect results were obtained for both $P_1$ and $P_2$ in both cases and then it decreases slowly to reach around 70\%  for the $P_4$. As for $M1$, the success rates deteriorated quickly reaching an unsatisfactory 50\% starting from the second-order $P_2$ of the parity benchmark due to the absence of nonlinear variations within this modulation range of the pump force. \\

\begin{figure*}
  \centering
  \includegraphics[width=1\linewidth]{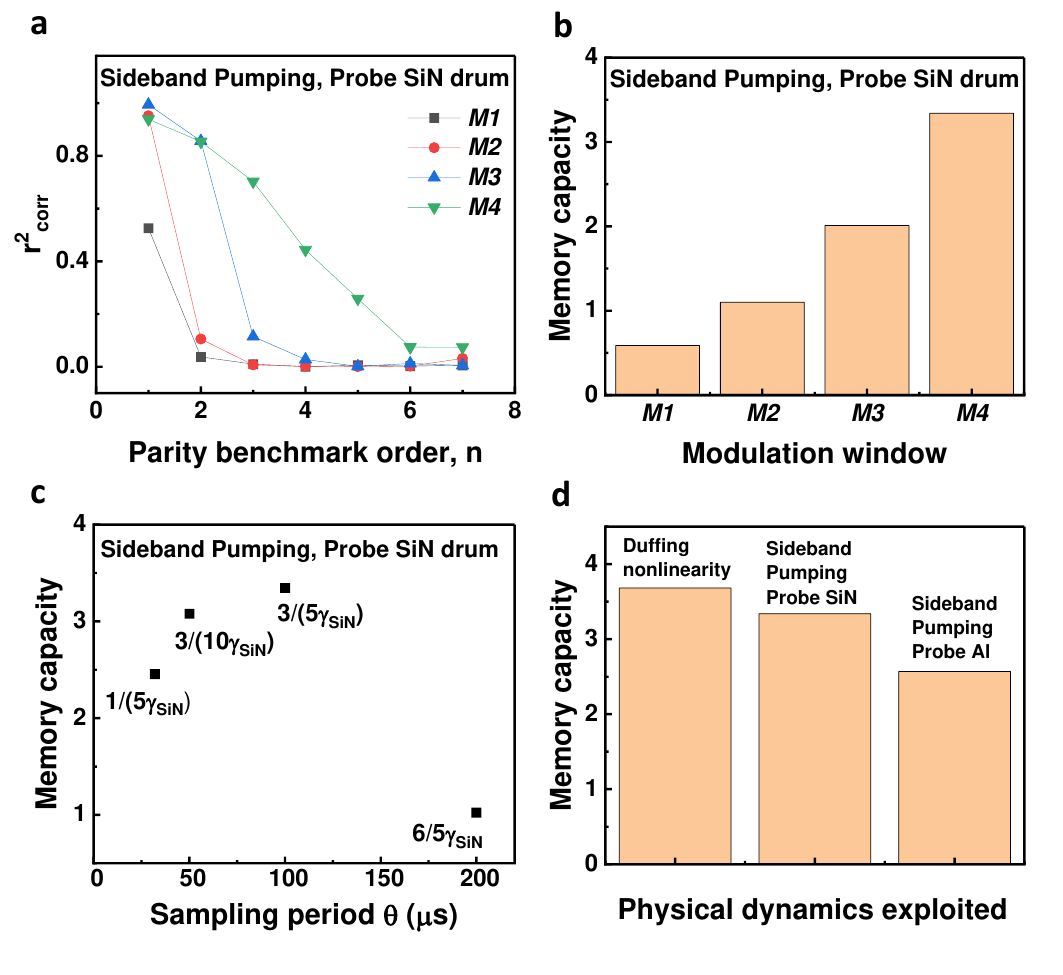}
  \caption{Evaluation of the short-term memory capacity. \textbf{a}: Square of the correlation coefficients for the first seven orders of the parity benchmark task, corresponding to different pump modulation window as shown in Fig. \ref{fig:parity_ProbSiN}\textbf{a}; \textbf{b}: The corresponding memory capacity; \textbf{c}: Memory capacity as a function of the sampling period $\theta$ and the damping rate $\gamma_{SiN}$; \textbf{d} Optimized memory capacity obtained in the three physical regimes investigated in this work: Duffing nonlinearity and the two blue-sideband pumping configurations.}
  \label{fig:STM}
\end{figure*}

\textbf{Short-term memory capacity } Short-term memory is the ability of a system to temporarily maintain a small amount of external input information in a transient state over a short duration \cite{MC_liao2023short}. Here, we evaluate the short-term memory capacity by calculating the square of the correlation coefficient between the predicted output $y$ and the target output $y_{target, n}$: 
\begin{equation}
    r_{corr}^2(n) = \frac{Cov(y,y_{target,n})^2}{Var(y)\times Var(y_{target,n})},
\end{equation}
Here, Cov, and Var are the covariance and variance, respectively, and the n corresponds to the order of the parity benchmark. The value of the memory capacity (MC) can be obtained by applying the following summation\cite{MC_liao2023short, furuta2018macromagnetic}:
\begin{equation}
    MC = \sum_{n=1}^{n_{max}} r_{corr}^2(n).
\end{equation}
The measured results from the parity benchmark tests in both the Duffing and the double-tone schemes are used to evaluate the MC, with $n_{max}$ = 7. Figure \ref{fig:STM}a-b show the evaluation results for the double-tone scheme when the SiN drum is probed and the phonon cavity is pumped with different amplitude modulation windows, as described in Fig.  \ref{fig:parity_ProbSiN}a. The optimal MC obtained is linked to $M4$ (MC = 3.345), followed by $M3$ (MC = 2.011), $M2$ (MC = 1.009) and lastly $M1$ (MC = 0.58). These results are in agreement with our previous analysis,  indicating that enhanced short-term memory performance occurs in the more nonlinear modulation regime induced by pump-mediated phonon interference. Furthermore, an analysis of the dependence of the MC on the sampling period $\theta$, which is correlated to the damping rate of the SiN resonator ($\gamma_{SiN}$), is presented in Fig. \ref{fig:STM}c. The corresponding modulation windows are chosen as $M4$ in order to maximize the effect of coupling via phonon interferences. The measurements show that the MC is highly sensitive to the values of $\theta$ and $\gamma_{1}$. The MC increases while increasing $\theta$ from $32\, \mu$s to $100\, \mu$s indicating successful reservoir computing predictions due to the high coupling between the virtual neurons of the reservoir. However, the effect of short-term memory then decreased as to reach approximately 1 for $\theta = 200\, \mu$s. On this timescale, the virtual nodes begin to experience a weakened transient regime as $\theta = 200 \,\mu$s approaches half of the mechanical damping time, leading to reduced coupling between virtual nodes and a deterioration of the short-term memory capacity. 

\subsection{The NARMA Benchmark}
The NARMA benchmark is widely used to evaluate nonlinear processing and fading memory in reservoir computing systems. The key feature of the NARMA benchmark is that the current output is generated through a nonlinear combination of many past inputs and outputs. As a result, it requires the reservoir that combines nonlinear processing with long fading memory \cite{barazani2020microfabricated}. A generalized version
of its input-output relationship for this benchmark is given by:
 \begin{equation}
     \hat{y}_n(k+1) =  0.3\hat{y}_n(k) + 0.05\hat{y}_n(k)\sum_{i=0}^{n-1}\hat{y}_n(k-i)+1.5u(k)u(k-n+1) + 0.1
 \end{equation}
where $n$ is a time-lag parameter, $k = \frac{t}{\tau}$ is the timestep.  The dimension of the input and output vectors corresponds to a total of $N$ data divided between $N_{tr}$ for the training phase and $N_{tst}$ that are utilized to test and evaluate the performance of the reservoir computing  experiment. The input $u(k)$ of the system consists of scalar random
numbers, drawn from a uniform distribution over the interval $[0 , 0.5]$. The $\hat{y}_n$ is the  output target. In order to evaluate the performance of the system based on the NARMA benchmark, we compute the Normalised Mean Squared Error (NMSE). Assuming that the target output vector is $\hat{y}$ and the predicted output vector by our reservoir computing is $y$, we adopt the following definition of the NMSE \cite{NARMA10, guo2024mems}:
\begin{equation}
\begin{split}
    MSE(\hat{y}_n,y_n) = \frac{1}{N_{tst}}\sum_{i=1}^{N_{tst}} (\hat{y}_n(i)-y_n(i))^2\\
    NMSE(\hat{y}_n,y_n) = \frac{MSE(\hat{y}_n,y_n)}{MSE(\hat{y}_n,mean(\hat{y}_n))}. 
    \end{split}
\end{equation}

As demonstrated in previous studies on the parity benchmark, the initial step in designing a reservoir computing framework is to find the optimal pump force modulation window. Here, we evaluated the system's NARMA performance across various modulation windows by probing the Al drum resonator, via a sideband pumping scheme, as presented in Fig. \ref{fig:parity_ProbAl}a. The resulting measurements for each modulation window are shown in Fig. \ref{fig:parity_ProbAl}d. The optimal NMSE values were achieved at the modulation window $M3$, coinciding with the nonlinear jump where phonon interference effects become dominant. This observation is consistent with our previous parity benchmark results, indicating that more complex nonlinear behavior benefits for mapping the input into higher-dimensional nonlinear spaces, thereby leading to the improved computing performance.\\

\begin{figure*}[htbp]
  \centering
  \includegraphics[width=1\linewidth]{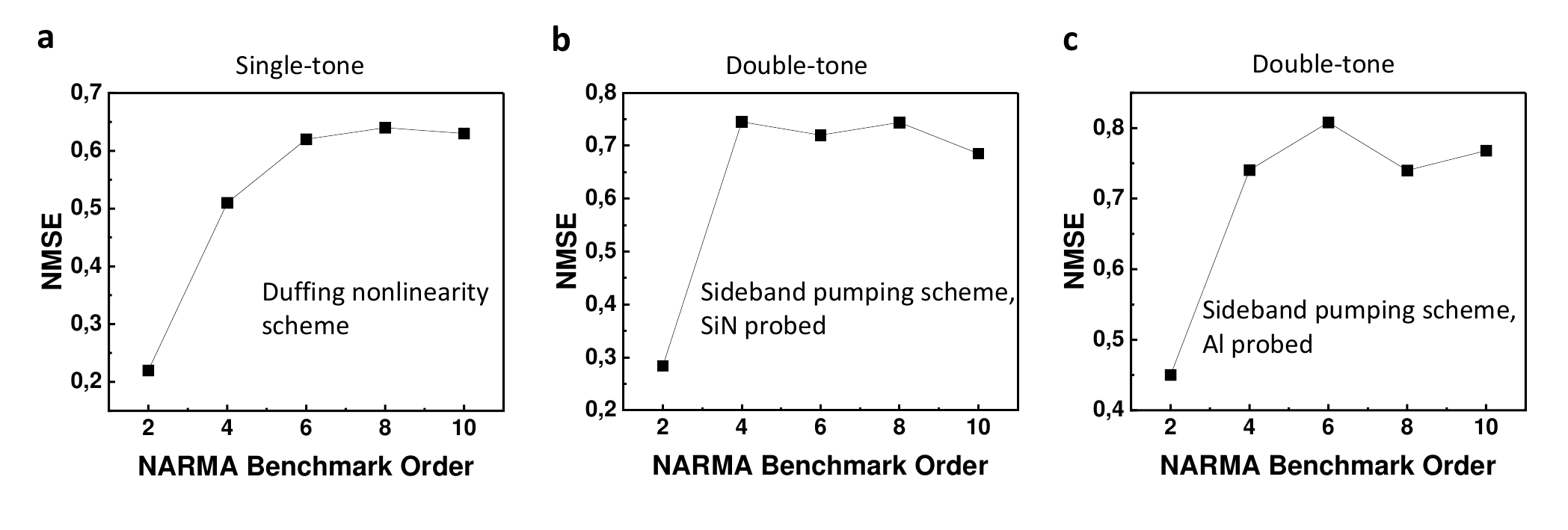}
  \caption{The obtained NMSE values as a function of the order of the NARMA Benchmark in the case of \textbf{a}: Duffing nonlinearity, and  blue sideband pumping the phonon-cavity while \textbf{b}: probing the SiN drum \textbf{c}: probing the Al drum.}
  \label{fig:NARMA_Benchmark}
\end{figure*}
By selecting an optimal modulation window that maximizes the system's nonlinear response, we evaluated various modulation schemes using the NARMA benchmark. The first approach leverages the Duffing nonlinearity, the predominant nonlinear characteristic of microelectromechanical systems, which has been successfully employed for reservoir computing in several prior studies \cite{Sylvestre, zheng2021parameters}. In our double-drum architecture, we exploit the Duffing properties of the SiN drum resonator. More details can be found in the SI part. Due to its circular geometry, the drum design yields a higher nonlinearity coefficient than conventional doubly-clamped beams \cite{cattiaux, venkatachalam2023effects}. The second and third methods are based on sideband pumping, which induces nonlinear phonon-transfer dynamics between two coupled vibrational modes, while probing the SiN phonon cavity and the Al resonator, respectively. The experimental findings are presented in Fig. \ref{fig:NARMA_Benchmark} and details regarding the training and testing parameters can be found in table \ref{tab:NARMA}. In the case of Duffing nonlinearity (Fig. \ref{fig:NARMA_Benchmark}a), the NMSE increases gradually from 0.2 and converges to approximately 0.6 at the tenth-order NARMA benchmark. The same trend is observed in the sideband pumping experiment while probing SiN (Fig. \ref{fig:NARMA_Benchmark}b) and Al drums (Fig. \ref{fig:NARMA_Benchmark}c). It is noticeable that the NMSE increases slightly while probing the SiN drum as it converges to approximately 0.7 and it reaches 0.8 for the NARMA10 in the case of probing the Al drum.  
\begin{table}[]
    \centering
\begin{tabular}{|l|c|c|c|}
\hline
 & \makecell{Duffing\\ nonlinearity}
 & \makecell{Blue sideband pump\\ probe phonon cavity}
 & \makecell{Blue sideband pump\\ probe Al resonator}\\
\hline
$\theta$ ($\mu$s) & 50 & 100 & 100 \\
\hline
$N_{\mathrm{tr}}$ & 1500 & 1500 & 1500 \\
\hline
$N_{\mathrm{tst}}$ & 400 & 400 & 400 \\
\hline
$\alpha_1$ & 0.51 & 0.52 & 0.36 \\
\hline
$\alpha_2$ & 0.85 & 0.35 & 0.51 \\
\hline
offset & 0.15 & 0.67 & 0.51 \\
\hline
\end{tabular}
    \caption{Reservoir computing parameters used for NARMA tests, corresponding to different nonlinear schemes.}
    \label{tab:NARMA}
\end{table}
\\

We compare the performance of this MHz range double-drum resonator with previously reported reservoir computing implementations using single MEMS resonators operating at kHz frequencies. For the parity benchmark, the success rates achieved in this double-drum resonator with the sideband pumping approach are slightly lower than those obtained using the Duffing nonlinearity alone. These results are comparable to those reported for MEMS accelerometer based reservoirs \cite{barazani2020microfabricated}, but remain below the performance achieved with doubly clamped beam MEMS implementations \cite{Sylvestre}. Regarding the NARMA benchmark, our NMSE results do not yet match the good performance of the kHz range MEMS, for instance without delay loops or mask functions \cite{guo2024mems, Sun2021, zheng2021parameters}. It should be noted that operation of mechanical resonators in the MHz frequency regime inherently could imposes stricter constraints on fading memory compared to kHz MEMS reservoirs when sufficiently high quality factors $Q$ are not achieved, which naturally affects performance on long-memory benchmarks such as NARMA. The fading memory of a mechanical reservoir relies on its decay time $T = 2Q/\Omega_m$. While operation in the MHz regime enables faster processing speeds, sufficiently high $Q$ values are required in order to maintain adequate memory capacity \cite{Sun2021}. For example, MEMS reservoirs operating at $\Omega_m/(2\pi)$ in a few hundreds of kHz range can naturally exhibit relatively long decay time $T\sim$ 100 $\mu$s with $Q\,\approx$ 100. A MHz resonator typically requires to have $Q>10^3$ for having a comparable level of memory capacity. In addition, the lower value of the $Q$ leads to increased effective noise due to the enlarged bandwidth, resulting in a reduced signal-to-noise ratio and weaker effective transduction between the driving force and mechanical displacement. In our experiment, the remaining performance gap is therefore primarily attributed to the limited memory depth and the reduced signal-to-noise ratio associated with the fast-decaying Al drum mode, which degrades the overall reservoir computing performance of the double-drum system. To address these limitations, further optimization of both device design and training strategies is required. On the device side, the mechanical properties of the top Al drum can be improved through nanofabrication optimization, such as reducing clamping losses through engineering the clamping surface (for instance, soft-clamping design) or increasing intrinsic stress by changing the Al deposition speed or thickness to enhance the quality factor. These improvements are expected to suppress noise and reduce the required driving strength in the two-tone sideband pumping scheme. On the algorithmic side, performance may be further enhanced by optimizing post-processing strategies or exploring alternative training approaches to extend the effective memory capacity \cite{guo2024mems, zhang2023survey}.\\

\section{Discussion}

This double-drum electromechanical resonator exhibits strong potential for high integration density and low driving energy. Because of its compact design, each vibrating element occupies an area of approximately 24 $\mu$m $\times$ 24 $\mu$m. And, the coupled-membrane architecture provides strong capacitive coupling ($C_{g0} \approx$ 9.61 fF), enabling efficient excitation of mechanical vibrations using a small $AC$ drive voltage. The effective driving voltages applied on  the device are $V_{p} = 326$ mV and $V_{d} = 0.6$ mV with a $DC$ bias of  $V_{dc}$ = 4 V. These values are obtained by taking into account the 20 dB attenuation of the driving signal, the 13 dB attenuation of the pump tone, and the voltage gain of the multiplier ($\times$ 10 in amplitude) applied to both the driving and pump tones in the input chain. The maximum power consumption to excite the double-drum device for creating the nonlinear operating regime is therefore estimated to be  $E_{el} = \frac{1}{2}C_{g0}V^2 =$ 12.53 fJ for one input (without interference) in the reservoir computing process, for the case of probing the SiN phonon cavity. To the best of our knowledge, this work demonstrates the lowest energy consumption per input (without interference) among reservoir computing implementations based on micromechanical resonators. We also evaluate the mechanical power consumption for the case of probing SiN drum in a sideband pumping scheme through $m_{Al, SiN} \Omega_{m(Al, SiN)}^3 X^2_{Al, SiN} /(2 Q_{Al, SiN})$, where $m$ is the effective mass of the drum resonator. Using the parameters $m_{SiN}=6.77\times 10^{-14}$ kg for the SiN drum, $m_{Al}=3.9\times 10^{-13}$ kg for Al drum. we consider the maximum mechanical displacements within the linear region $X_{SiN}\approx$ 5 nm for the SiN drum and take $X_{Al}\approx$ 500 pm which is about an order of magnitude smaller than $X_{SiN}$ in the two-tone driving scheme when probing SiN drum \cite{alok}. The estimated mechanical power consumption is $P_{mech}$ = 14.88 pW and 5.87 pW, for SiN and Al drum respectively. The double-drum electromechanical system requires 40 ms to map all masks of the one input data, resulting in an energy dissipation of 8.3 pJ per input. For processing a total of $N_{tr}$ and $N_{tst}$ inputs in this work, the double-drum resonator system exhibits a mechanical power consumption of approximately 15.7 nW.\\

The processing speed for MEMS-based reservoir computing is primarily limited by the decay time of the mechanical response, given by $T_{SiN, Al}= 2Q_{SiN, Al}/\Omega_{SiN, Al}$. The delay time of the feedback loop, $\tau$ = $N_{mask}\times \theta$ determines the data processing rate of the system. To maintain operation in the transient regime and enable effective coupling between virtual nodes, the virtual node separation should satisfy $\theta$, where $T$ is the decay time of the physical system. In this work, the optimal value of $\theta$ for reservoir computing is 100 $\mu$s with $N_{mask}=400$ in the double-tone scheme based on two coupled drum resonators, corresponding to a processing rate of 1/$\tau$ = 25 Hz. To increase the processing speed, the ideal approach is to operate MEMS resonators at higher resonance frequencies while maintaining a lower quality factor. However, an optimal balance between resonance frequency and quality factor is required: excessively low $Q$ values increase mechanical noise and necessitate higher drive amplitudes, leading to increased energy consumptions. In the present double-drum resonator having resonance frequencies in MHz range, such a balance is required. For instance, the SiN drum operates in the $>$ 10 MHz frequency range with a suitable $Q \sim 10^4$, giving the decay time $T_{SiN}\approx$ 330 $\mu$s. For characterizing Duffing nonlinearity based reservoir computing, the best success rates are obtained around $\theta$ = 50 $\mu s$, having $\theta \approx T_{SiN}/6$ for SiN drum. Under these conditions, approximately six virtual neurons are generated within one oscillation period, which is a reasonable value compared with previous reports \cite{Feedbback}. \\

%
%
In the coupled mechanical resonators based double-tone scheme, whether probing the phonon cavity or the Al drum, we obtain an optimal virtual node separation $\theta$ = 100 $\mu$s for both the parity and NARMA benchmarks. In this regime, the $\theta \gg T_{Al}$ = 12.5 $\mu$s while $\theta < T_{SiN}$. In the delayed feedback system, it is generally expected that when the $\theta$ much larger than the decay time $T$ of the nonlinear node, temporal coupling between virtual nodes is lost, causing each node to behave as an effectively self-coupled unit and reducing the diversity of reservoir states. Our experimental results do not contradict this principle. In the present double-drum system, the nonlinear dynamics exploited for reservoir computing processing arise from the energy (in the form of vibration phonons) coherent transfer between two coupled modes. Consequently, although one resonator exhibits a fast decay, the presence of the slower resonator sustains the system in a transient dynamical regime, thereby preserving effective temporal coupling and reservoir state diversity. The coupled resonator design effectively extends the reservoir memory through multimode coupling, even when the virtual node separation exceeds the decay time of the faster resonator. \\

Within the phonon-cavity framework, when the cavity is pumped at its blue sideband, a cavity force is generated that reduces the effective damping rate of the coupled resonators through modulating the mechanical susceptibility. This phenomenon, known as the optical damping effect, has been observed in both coupled mechanical resonators and optomechanical systems \cite{sun2016correlated, alok, PhysRevApplied_XZ_2019, zhou2021electric}. The effective damping rate can be expressed as $\gamma_{eff}$ = $\gamma_m$ - $\gamma_{opt}$, where the $\gamma_m$ is the intrinsic damping rate in the absence of the pump tone and $\gamma_{opt}$ represents the pump-induced modification \cite{alok}. Based on Eq.~\ref{eq:motion} and Eq.~\ref{eq:x1x2}, the optically induced damping terms can be expressed as functions of an effective coupling strength $g^2$ and the pump frequency $\Omega_p$, $\gamma_{opt,1} \approx \frac{g^2\gamma_2}{(\Omega_2 - (\Omega_p-\Omega_1))^2+\frac{\gamma_2^2}{4}}$ and  $\gamma_{opt,2} \approx\frac{g^2\gamma_1}{(\Omega_2 - (\Omega_p-\Omega_1))^2+\frac{\gamma_1^2}{4}}$. Using the experimental parameters and considering the pump amplitude corresponding to modulation window M4 in \ref{fig:parity_ProbSiN}a, the decay time of the SiN drum varies between 383 $\mu$s and 434 $\mu$s. In our RC experiment, the implementation of a virtual node separation of $\theta$ = 100 $\mu$s for both the parity and NARMA benchmarks ensures that the system operates in the transient regime. Therefore, when exploiting the two-tone driving scheme for reservoir computing, the pump amplitude must be carefully controlled to avoid significant changes in the mechanical decay time, which would lead to a loss of the benefits of the transient regime. \\

In summary, this work presents a novel reservoir computing scheme based on a double-drum resonator system that leverages concepts from phonon-cavity electromechanics. In the two-tone scheme, controllable nonlinear phonon transfer dynamics between two coupled resonators are induced by applying a pump tone at the blue sideband of the phonon-cavity while probing one of the coupled resonators. The pump tone acts as a coherent data bus, enabling the masked input data to be mapped onto both coupled drums for reservoir computing. Although further efforts are still required to improve reservoir computing performance, this sideband pumping approach provides a way for extending a single MEMS based reservoir computing to multimode systems. In this double-drum design, one of the coupled modes to function as a dedicated sensor, while the detected signals are transduced into mechanical properties, for instance the variations of phonon numbers or resonance frequency, and are coherently transferred to the second resonator for reservoir computing. This separation of sensing and processing within a single electromechanical platform advances the development of compact MEMS systems that integrate sensing and computing functionalities. Finally, we would like to emphasize that the proposed reservoir computing scheme based on the sideband pumping technique is not limited to our double-drum resonator architecture and does not require the resonance frequencies of the two coupled resonators to be closely matched. Compared with conventional mechanical coupling schemes, this method enables coupling between two spatially separated resonators with distinct resonance frequencies, without requiring high nanofabrication precision to ensure closely matched resonances. It can be straightforwardly extended to other multimode coupling platforms, including mechanically coupled resonator arrays and optomechanical systems, providing additional flexibility for device design and system integration.
\section{Materials and methods}
\label{MM}
\textbf{Device fabrication:} The double-drum electromechanical resonator, measured in this work, consists of two suspended membrane. The device fabrication process begins with a high-resistivity silicon substrate ($>$10k $\Omega \cdot$cm.) covered with a stoichiometric SiN thin film (90 nm in thickness), having $\sim$1 GPa tensile stress. To release the SiN membranes from the substrate, circularly symmetric holes with a diameter of 350 nm are first patterned in the SiN layer using electron-beam lithography. A reactive ion etching (RIE) process employing $SF_6$ and Ar gases is then used to remove the exposed SiN through those patterned holes on the resist. This step is followed by selective isotropic etching of the underlying silicon substrate using $XeF_2$ (Xenon difluoride), which exhibits a high selectivity between silicon and SiN. This two-step etching process releases the SiN membranes while preserving fully clamped boundary conditions at the edges, forming suspended drum resonators. To form capacitive coupling scheme, about 20 nm Al thin film is deposited on the SiN drum as a conductive layer. The suspended top gate of the SiN drum resonator is fabricated using a polymethyl methacrylate (PMMA) resist layer with a thickness of approximately 350 $\pm$ 50 nm which serves as a sacrificial support layer. The PMMA is defined through a soft-bake and reflow process \cite{xin_zhou, xu2023fabrication}. Subsequently, the top-gate pattern is defined in a second layer of methyl methacrylate (MMA) resist deposited on top of the PMMA layer. The fabrication process is completed by depositing an approximately 550 nm thick Al film, followed by a lift-off process. During lift-off, the sacrificial PMMA support layer is also removed, resulting in a suspended top-gate structure. The SiN drum has a diameter $\approx$ 22 $\mu$m and the Al drum has a diameter of $\approx$ 24 $\mu$m. The separation between the two drums is $\approx$ 
350 nm. \\

\textbf{Implementation of measuring equipment:} The microwave source used for microwave interferometry measurements is an Rohde $\&$ Schwarz SMB100B signal generator, which provides ultra-low phase noise. The microwave power shining to the double-drum through transmission line is around -5 dBm. A Yokogawa GS200 source is used to supply the $DC$ bias voltage $V_{dc}$. The $AC$ drive signals are generated using a Keysight Technologies 33600A arbitrary waveform generator for the pump tone and a Zurich Instruments UHFLI 600 MHz lock-in amplifier for the probe tone. After frequency down-conversion of the microwave signal to the MHz range, the mechanical displacement excited by the probe tone is detected by the lock-in amplifier as a voltage signal. This signal is amplified by a scaling factor $gl$ by the lock-in amplifier and is then fed into an AMD Spartan-7 FPGA (XC7S100) to introduce a time delay $\tau$. It is then added with the the masked data (mask $m(t)$ onto the input data $u(t)$) for the output of the FPGA. An AD835 analog multiplier is used to multiply the incident pump signal with the FPGA output, in order to modulate the pump force applied to the double-drum resonator. In the measurement, the output of this multiplier has been calibrated in order to control the pump force modulation range and to maintain linearity. \\

\textbf{Training:} The displacement vector, created from the force amplitude modulation,  will be linearly combined and give the final output $y(t)$ based on the formula: 
\begin{equation}
    y(t) = W^{T}x(t) .
\end{equation}
The weight vector $W$ will be trained in a supervised training scheme as to provide a prediction $y(t)$ as close as possible to the target $y_{target}(t)$. The training is done to have the minimizing the mean squared error between $y(t)$ and $y_{target}(t)$ by applying a ridge regression as following:
\begin{equation}
    W = y_{target} Z^T(ZZ^T+\lambda I)^{-1}
\end{equation}
The $Z$ is the data displacement matrix consisting of $x(t)$ and $\lambda$ is the regularization parameter. When the training of the weight vector $W$ ends, we will test the performance of our system for different inputs \cite{Brunner_ridge_regression, Sylvestre}. All computing processes are based on 2000 input bits. The first 100 inputs are discarded to eliminate offset effects from previous measurements. The subsequent 1500 inputs are used for training, and the remaining 400 inputs are used for prediction.

\bibliography{sample}

\section*{Acknowledgements (not compulsory)}
The authors would like to acknowledge financial support from the French National Research Agency, ANR-MORETOME, No. ANR-22-CE24-0020-01, and Chist-ERA NOEMIA project with contract ANR-22-CHR4-0006-01. This work was partly supported by the French Renatech network. X.Z. also would like to thank Toky-Harrison Rabenimanana for his contributions in developing the reservoir computing setup.

\section*{Author contributions statement}
X.Z. conceived the experiments. L.F. and X.Z. fabricated the samples in the clean room. P.L. and X.Z. developed the experimental setup. T.F., L.F., and X.Z. conducted the experiments and analyzed the results. T.F. and X.Z. wrote the manuscript, L.F.and B.D-R reviewed the manuscript, G-E.C., J-Y.O., Y.T., Y.P. contributed to the discussions. X.Z. supervised the project and acquired the research funding.

\end{document}